\documentclass[11pt]{article}

\usepackage{acl}

\usepackage{times}
\usepackage{latexsym}
\usepackage{booktabs}
\usepackage{makecell}
\usepackage{setspace}
\usepackage[T1]{fontenc}

\usepackage[utf8]{inputenc}
\usepackage{enumitem}
\usepackage{microtype}
\usepackage{xcolor}
\usepackage{inconsolata}

\usepackage{graphicx}
\usepackage{enumitem}
\usepackage{xcolor}
\usepackage{multirow}
\usepackage{booktabs}
\usepackage{makecell}
\usepackage{multirow}
\usepackage{graphicx}
\usepackage{siunitx}

\newcommand{\mh}[1]{{\scriptsize\textbf{#1}}}

\usepackage[most]{tcolorbox}
\newtcolorbox{resultbox}{colback=blue!5!white, colframe=blue!50!black,
  boxrule=0.4pt, arc=2pt, left=6pt, right=6pt, top=3pt, bottom=3pt}

\title{\textsc{UK-PRBench}: A Paragraph-Level Precedent Retrieval Benchmark \\ for United Kingdom Case Law}

\author{Damith Premasiri \\
Lancaster University, UK \\
  \texttt{d.dolamullage@lancaster.ac.uk} \\\And
  Tharindu Ranasinghe \\
  Lancaster University, UK \\
  \texttt{t.ranasinghe@lancaster.ac.uk} \\}

\begin{document}
\maketitle
\begin{abstract}
Prior case retrieval (PCR) aims to identify precedent cases relevant to a given query case. Existing PCR benchmarks and methods predominantly operate at the document level, treating entire judgments as the unit of relevance. This formulation is suboptimal for legal practitioners, as judgments address multiple legal issues and only a small subset of paragraphs is relevant to a particular query. Addressing this gap, we introduce \textsc{UK-PRBench}, a benchmark for paragraph-level precedent retrieval in UK case law, constructed from judgments obtained from the UK National Archives and covering a broad range of UK courts and tribunals. Furthermore, we evaluate state-of-the-art retrieval models and establish baseline results. Our experiments show that paragraph-level precedent retrieval remains challenging for current retrieval approaches, highlighting substantial room for improvement. \textsc{UK-PRBench} provides a standardised benchmark for evaluating fine-grained precedent retrieval and advancing retrieval systems for the UK legal domain.
\end{abstract}

\section{Introduction}
Prior case retrieval (PCR) is the task of identifying relevant precedent cases for a given query case \cite{ijcai2020p484, feng-etal-2024-legal}. It is a foundational application of legal natural language processing \cite{10.1145/3626772.3657650}, supporting legal practitioners in citing related precedents to establish applicable law and construct arguments, as courts are generally bound to decide cases consistently with earlier rulings \cite{joshi-etal-2023-u}. As the volume of case law grows, manually navigating available case law for relevant precedents becomes increasingly impractical \cite{deng-etal-2024-learning}, motivating the development of automated retrieval systems. Consequently, the development of automated PCR systems has attracted significant attention from both the legal informatics and information retrieval communities \cite{hou-etal-2025-clerc}.

The development of PCR systems has been facilitated by benchmark datasets covering a range of jurisdictions and retrieval settings. Existing benchmarks include \textsc{COLIEE} \cite{10.1145/3769126.3785016} and \textsc{IL-PCR} \cite{joshi-etal-2023-u} for English-language case law from Canada and India, and \textsc{CAIL-SCM} \cite{10.1145/3735127} and \textsc{LeCaRD} \cite{10.1145/3404835.3463250} for Chinese case law. Machine learning (ML) approaches for PCR have also progressed from expert-knowledge and citation-network methods \cite{10.1007/s10506-009-9075-y} to domain-specific pre-trained language models such as \texttt{Legal-BERT} \cite{chalkidis-etal-2020-legal}, and more recently to dense embedding-based retrieval \cite{joshi-etal-2023-u}.

However, existing datasets and methods operate at the document level, treating an entire judgment as the unit of relevance \cite{paul-etal-2025-il}. This formulation does not fully reflect the structure of judicial decisions. A judgment may address multiple legal issues, factual circumstances, and holdings \cite{10.1145/3583780.3615125}, meaning that only a small number of paragraphs may be relevant to a particular query case, while much of the remaining judgment may be unrelated \cite{10.1007/978-3-031-29168-5_7}. Document-level retrieval can therefore identify a relevant judgment while still requiring practitioners to locate the specific passages that support the legal issue under consideration. This can reduce retrieval precision and increase the amount of irrelevant material that must be examined \cite{10.1007/978-3-030-99736-6_2}. 

Paragraph-level retrieval addresses this limitation by directly identifying the relevant portions of precedent cases, providing a more fine-grained formulation of PCR and a closer approximation of the information retrieval process involved in legal research \cite{ijcai2020p484}. Despite its practical importance, paragraph-level precedent retrieval remains comparatively underexplored, particularly for UK case law.  The absence of such a benchmark limits systematic evaluation of retrieval models in the UK legal domain.

 To address these gaps, we introduce \textsc{UK-PRBench}, a benchmark for paragraph-level precedent retrieval in UK case law. The benchmark is constructed from cases obtained from the UK National Archives and covers judgments from a broad range of UK courts and tribunals. Using \textsc{UK-PRBench}, we establish baseline results for state-of-the-art transformer and embedding-based retrieval models. Our results show that paragraph-level precedent retrieval remains challenging for current retrieval methods, highlighting the need for models and retrieval strategies specifically designed to identify relevant passages.

The \textbf{main contributions} of this paper are,
\begin{enumerate}[label={(\textcolor{blue}{\textit{\roman*}})}, wide, labelindent=0pt]
\setlength{\itemsep}{-2pt}
    \item We introduce \textsc{UK-PRBench} \footnote{Available at \url{https://github.com/DamithDR/uk-case-retrieval}}, a benchmark for paragraph-level precedent retrieval over UK case law, constructed from a large collection of judgments spanning multiple courts and tribunals.
    \item We evaluate a range of neural and embedding-based retrieval models,
    establishing baseline results for paragraph-level legal information retrieval.
\end{enumerate}

\section{Related work}

\subsection{Document-level Case Retrieval}
Case retrieval is primarily studied as a document-level task, in which the goal is to retrieve relevant precedents from a large pool of cases given a query case \cite{feng-etal-2024-legal}. Case retrieval is important across every legal system, and datasets have been constructed for a range of jurisdictions, including Chinese \cite{10.1145/3404835.3463250, 10.1145/3583780.3615125}, Indian \cite{joshi-etal-2023-u, mandal2017overview}, German \cite{wrzalik-krechel-2021-gerdalir}, and United States \cite{mahari-etal-2024-lepard} case law. More recently, \citet{kim-etal-2025-legalsearchlm} introduced LEGAR-BENCH, a large-scale Korean benchmark covering 411 crime types over 1.2M candidate cases, together with LegalSearchLM, a retrieval model that performs legal-element reasoning over the query and generates grounded content via constrained decoding.

A substantial line of work develops retrieval methods tailored to the legal domain. Pre-training objectives designed for legal case retrieval have been proposed \cite{10.1145/3735127}, including SAILER, a structure-aware pre-trained model that exploits the organisation of judgments \cite{10.1145/3539618.3591761}. Others incorporate legal semantics more explicitly: \citet{https://doi.org/10.1155/2022/2511147} combine topic distributions and legal-entity facts with BERT-based paragraph aggregation to better represent long case documents, while \citet{Dan2025} propose the Legal Event-Context Model (LECM), which integrates legal events with their surrounding context through an attention mechanism to improve accuracy and interpretability. To address data scarcity, \citet{Bui2024} use large language models to augment training data for case entailment and retrieval. More recently, \cite{premasiri-etal-2025-llm} experimented with LLM-based encoders \cite{ranasinghe-etal-2025-musts} and showed that LLMs struggle with the task. 

\subsection{Paragraph-level Retrieval}
A smaller body of work moves beyond document retrieval towards fine-grained, paragraph-level retrieval. \citet{tang-clematide-2021-searching} build a Swiss (German) dataset of cases and statutes, extract case-to-statute citations, and estimate paragraph-level semantic similarity using a link-based method, showing that neural similarity modelling can be improved with an extended attention mask that suppresses input noise. On judgments from the European Court of Human Rights (ECtHR), \citet{t-y-s-s-etal-2024-query} extract query-relevant paragraphs by aligning them with guideline themes and descriptions, and \citet{upadhya-t-y-s-s-2025-lexclipr} address cross-lingual paragraph retrieval. Query reformulation has also been explored, as in GuRE, a generative query rewriter for legal passage retrieval \cite{kim-etal-2025-gure}. These efforts demonstrate the value of finer retrieval granularity, but they target other jurisdictions and settings. As a result, paragraph-level precedent retrieval for UK case law remains unaddressed, which is the gap this paper addresses.

\subsection{Legal NLP Resources in the UK}
As noted by recent studies such as \citet{Premasiri2025-survay}, there remains a considerable lack of publicly available datasets and benchmarks for legal NLP in the UK. Existing resources are largely limited to a small number of tasks, including judgment prediction \cite{10.1007/978-3-032-32023-0_23} and topic classification \cite{chalkidis-sogaard-2022-improved}, while many other legal NLP tasks remain comparatively underexplored. This gap is also evident in widely used English legal NLP benchmarks such as LexGLUE \cite{{chalkidis-etal-2022-lexglue}}, which do not include any datasets representing the UK legal system. As a result, we believe that the dataset introduced in this paper will contribute to the growing legal NLP ecosystem in the UK.

\section{Constructing \textsc{UK-PRBench}}

\begin{table}[h]
	\centering
	\small
	\renewcommand{\arraystretch}{1.0}
	\begin{tabular}{@{} l c @{}}
		\toprule
		\textbf{Court} & \textbf{Coverage} \\
		\midrule
		\multicolumn{2}{@{}l}{\textit{\textbf{From Specific Courts}}} \\
		United Kingdom Supreme Court            & 2009--2026 \\
		Privy Council                           & 2009--2026 \\[2pt]
		\multicolumn{2}{@{}l}{\textit{\textbf{Court of Appeal}}} \\
		\quad Civil Division                    & 2001--2026 \\
		\quad Criminal Division                 & 2003--2026 \\[2pt]
		\multicolumn{2}{@{}l}{\textit{\textbf{High Court (England and Wales)}}} \\
		\quad Administrative Court              & 2003--2026 \\
		\quad Admiralty Court                   & 2003--2026 \\
		\quad Chancery Division                 & 2003--2026 \\
		\quad Commercial Court                  & 2003--2026 \\
		\quad Family Division                   & 2003--2026 \\
		\quad Intellectual Property Enterprise Court & 2013--2026 \\
		\quad King's / Queen's Bench Division   & 2003--2026 \\
		\quad Mercantile Court                  & 2008--2014 \\
		\quad Patents Court                     & 2003--2026 \\
		\quad Senior Courts Costs Office        & 2003--2026 \\
		\quad Technology and Construction Court & 2003--2026 \\[2pt]
		\multicolumn{2}{@{}l}{\textit{\textbf{Other Courts}}} \\
		\quad Crown Court                       & 2020--2026 \\
		\quad County Court                      & 2019--2026 \\
		\quad Family Court                      & 2014--2026 \\
		\quad Court of Protection               & 2009--2026 \\
		\bottomrule
	\end{tabular}
    \caption{Temporal coverage of judgments from selected UK courts in the UK National Archives Case Law collection.}
    \label{tab:uk-courts}
    
\end{table}

\begin{table}[h]
	\centering
	\small
	\renewcommand{\arraystretch}{1.0}
	\begin{tabular}{@{} l c @{}}
		\toprule
		\textbf{Tribunal} & \textbf{Coverage} \\
		\midrule
		\multicolumn{2}{@{}l}{\textit{\textbf{From Specific Tribunals}}} \\
		Investigatory Powers Tribunal                      & 2023--2024 \\
		Special Immigration Appeals Commission             & 2003--2026 \\
		Employment Appeal Tribunal                         & 2021--2026 \\[2pt]
		\multicolumn{2}{@{}l}{\textit{\textbf{Upper Tribunals}}} \\
		\quad Administrative Appeals Chamber               & 2005--2026 \\
		\quad Immigration and Asylum Chamber               & 2007--2026 \\
		\quad Lands Chamber                                & 2014--2026 \\
		\quad Tax and Chancery Chamber                     & 2016--2026 \\[2pt]
		\multicolumn{2}{@{}l}{\textit{\textbf{First-tier Tribunals}}} \\
		\quad Care Standards                               & 2024--2026 \\
		\quad General Regulatory Chamber                   & 2009--2026 \\
		\quad Tax Chamber                                  & 2019--2026 \\
		\quad Land Registration Division (Property Chamber) & 2024--2026 \\
		\quad Primary Health Lists                         & 2025--2026 \\[2pt]
		\multicolumn{2}{@{}l}{\textit{\textbf{Historic Tribunals}}} \\
		\quad Immigration Services Tribunal                & 2001--2009 \\
		\quad Consumer Credit Appeals Tribunal             & 2008--2013 \\
		\quad Estate Agents Tribunal                       & 2010 \\
		\quad Claims Management Services Tribunal          & 2010--2011 \\
		\quad Transport Tribunal                           & 2000--2010 \\
		\bottomrule
	\end{tabular}
    
    \caption{Temporal coverage of judgments from selected UK tribunals in the UK National Archives Case Law collection.}
    \label{tab:uk-tribunals}
\end{table}

\subsection{Data source}

We construct \textsc{UK-PRBench} from UK judgments available through the UK National Archives Case Law service.\footnote{Available at \url{https://caselaw.nationalarchives.gov.uk/}} The service provides judgments from a wide range of UK courts and tribunals, including the Supreme Court, Court of Appeal, High Court, and various specialist courts and tribunals. Table \ref{tab:uk-courts} and Table \ref{tab:uk-tribunals} summarise the courts and tribunals represented in the source collection and their available coverage.

We selected the National Archives as the primary data source because it provides case documents in machine-readable formats and covers a broad range of UK courts and tribunals. Although other legal repositories, such as BAILII,\footnote{Available at \url{https://www.bailii.org/form/search_cases.html}} provide extensive collections of UK case law, their terms of access do not support the programmatic acquisition required for constructing our benchmark. We therefore obtained a transactional licence from the UK National Archives to access and use the case data for this research.

The National Archives provides judgments in several formats, including XML, HTML, and plain text. We use the XML representation because it provides structured access to case metadata and judgment text. From each judgment, we extract its neutral citation, judgment date, and sequence of paragraphs. We additionally identify references to previously decided cases and the paragraphs cited within those cases. These citation relationships form the basis for constructing the paragraph-level retrieval instances in \textsc{UK-PRBench}.

\subsection{Citation Extraction and Instance Construction}

\begin{figure}[h]
	\centering
	\includegraphics[width=\columnwidth]{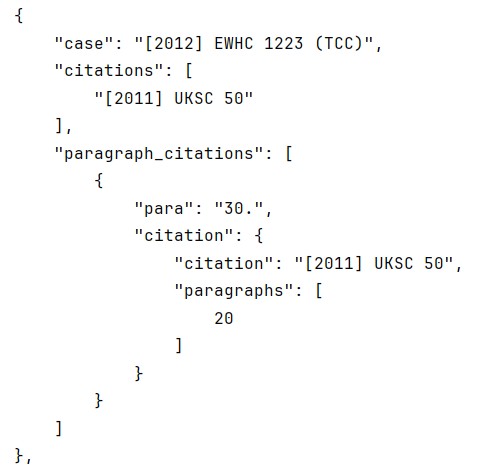}
	\caption{Example of case citations}
	\label{fig:citations}
\end{figure}

We identify case citations and cited paragraph numbers using a collection of regular-expression patterns designed to capture the citation formats occurring in UK judgments. For each identified citation, we determine whether the cited case is available in our collection and, where applicable, identify the specific paragraph or paragraphs referenced by the citation. Figure \ref{fig:citations} illustrates an example of the citation information extracted from a judgment.

To prevent models from exploiting explicit citation information rather than learning the semantic relationship between the query and precedent paragraphs, we replace extracted case citations with placeholders before constructing the retrieval inputs. The original citation information is retained separately in JSON format and is used to establish the corresponding query--precedent paragraph relationships.

We subsequently remove citation instances for which the cited source case is unavailable in our collection. This is necessary because some citations refer to judgments that fall outside the temporal or institutional coverage of the National Archives collection. For example, the National Archives collection does not contain Supreme Court judgments from before 2009. We also exclude citations to European courts, including the European Court of Human Rights and the Court of Justice, because these judgments were not collected as part of our dataset.

\subsection{Benchmark Construction}

The resulting collection contains $64,284$ judgments. From the citation relationships identified in these judgments, we construct paragraph-level retrieval instances consisting of an anchor paragraph, a relevant precedent paragraph, and candidate paragraphs. The anchor paragraph serves as the query, while the cited paragraph in the referenced case constitutes the positive retrieval target.

 We also include hard negative paragraphs drawn from the same precedent case. These negative examples are particularly important for paragraph-level retrieval because paragraphs within the same judgment may concern related legal issues while only one paragraph is explicitly cited by the query.

 We construct two benchmark configurations.

 \begin{itemize}
    \item \textbf{Single Paragraph Retrieval (SP):} Uses the citing paragraph as the query and the cited paragraph as the positive target.

    \item \textbf{Contextual Paragraph Retrieval (CP):} Extends the query and target with their immediate local context by incorporating the preceding and following paragraphs. This configuration evaluates whether contextual information improves retrieval when the relevant legal passage is not considered in isolation.
\end{itemize}

\subsection{Data Splits}

We divide the resulting instances into training, development, and test sets. The training set contains 15,363 anchor instances, the development set contains 2,128 instances, and the test set contains 4,403 instances. The candidate pool contains 8,739 paragraphs, including negative candidates designed to increase the difficulty of the retrieval task.

\begin{figure}[h]
	\centering
	\includegraphics[width=0.9\columnwidth]{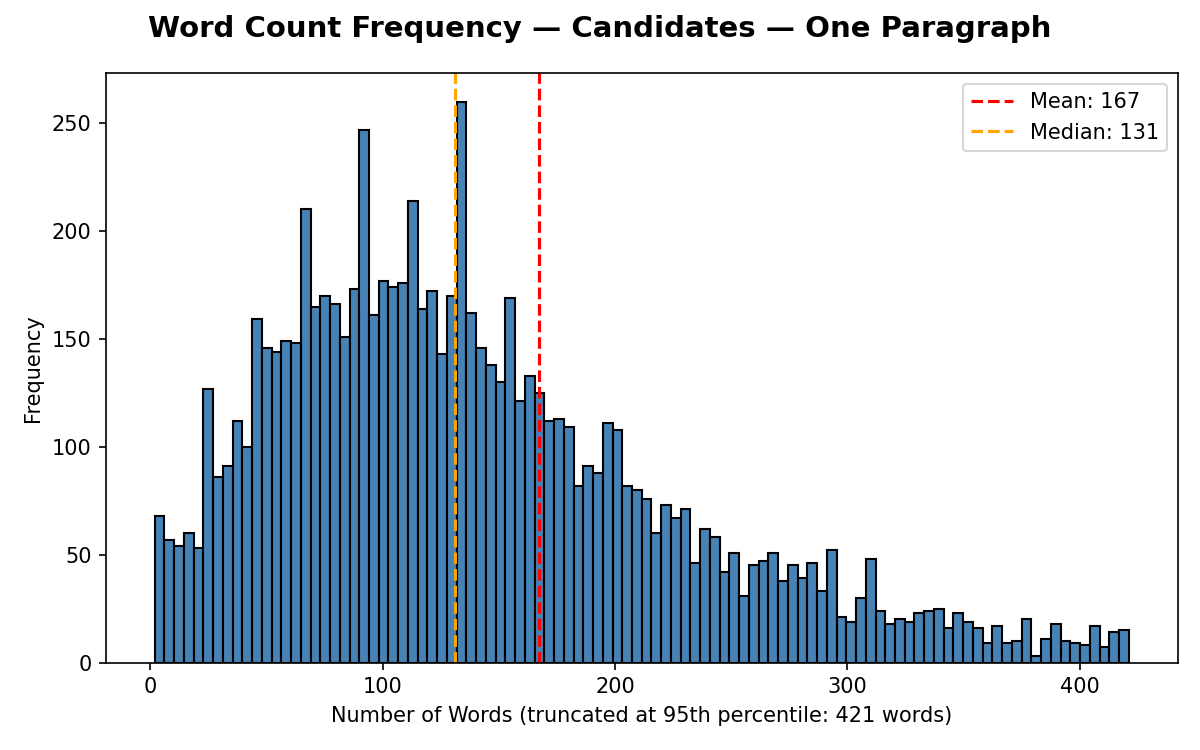}
	\caption{Word frequency graph for single paragraph candidates}
	\label{fig:1p_candidates}
\end{figure}

\begin{figure}[h]
	\centering
	\includegraphics[width=0.9\columnwidth]{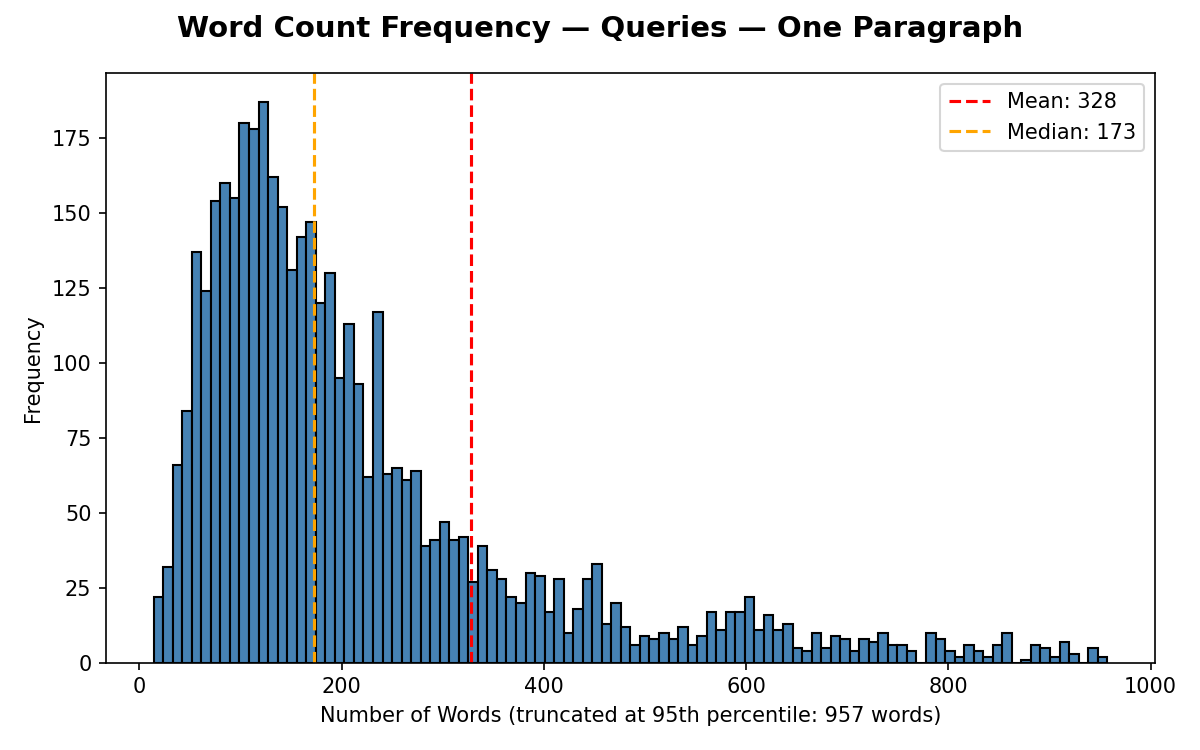}
	\caption{Word frequency graph for single paragraph queries}
	\label{fig:1p_queries}
\end{figure}

\begin{figure}[h]
	\centering
	\includegraphics[width=0.9\columnwidth]{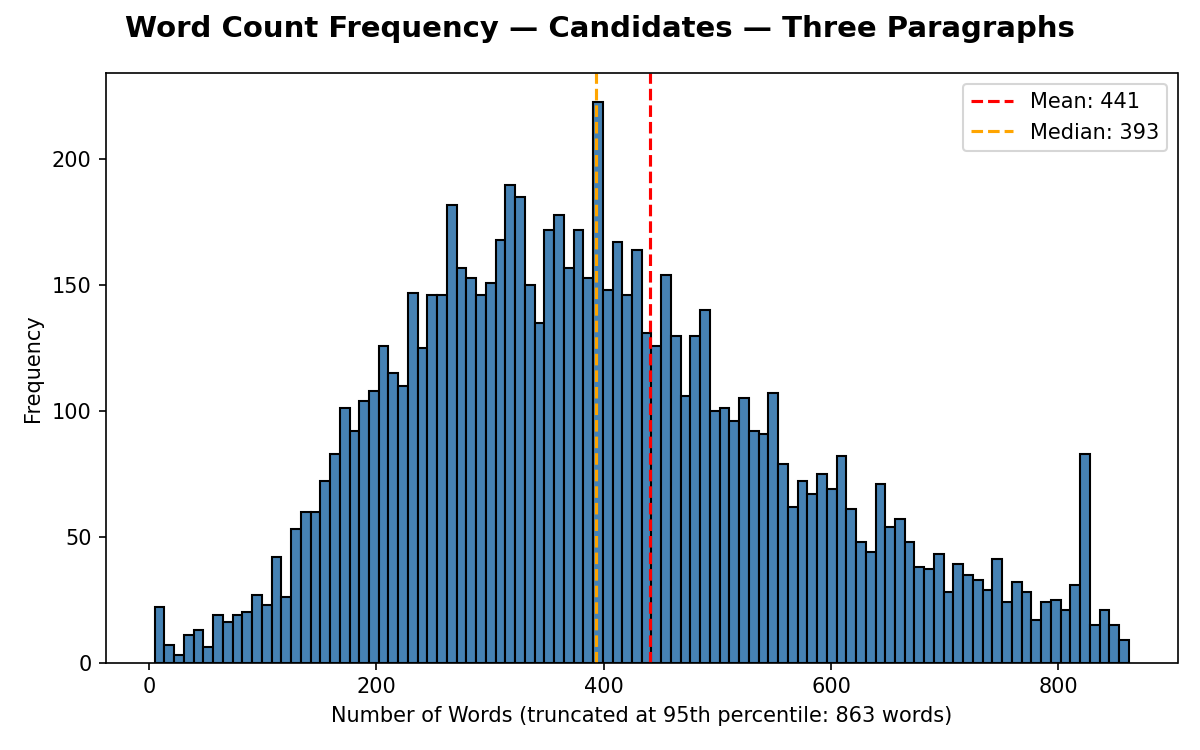}
	\caption{Word frequency graph for three paragraph candidates}
	\label{fig:3p_candidates}
\end{figure}

\begin{figure}[h]
	\centering
	\includegraphics[width=0.9\columnwidth]{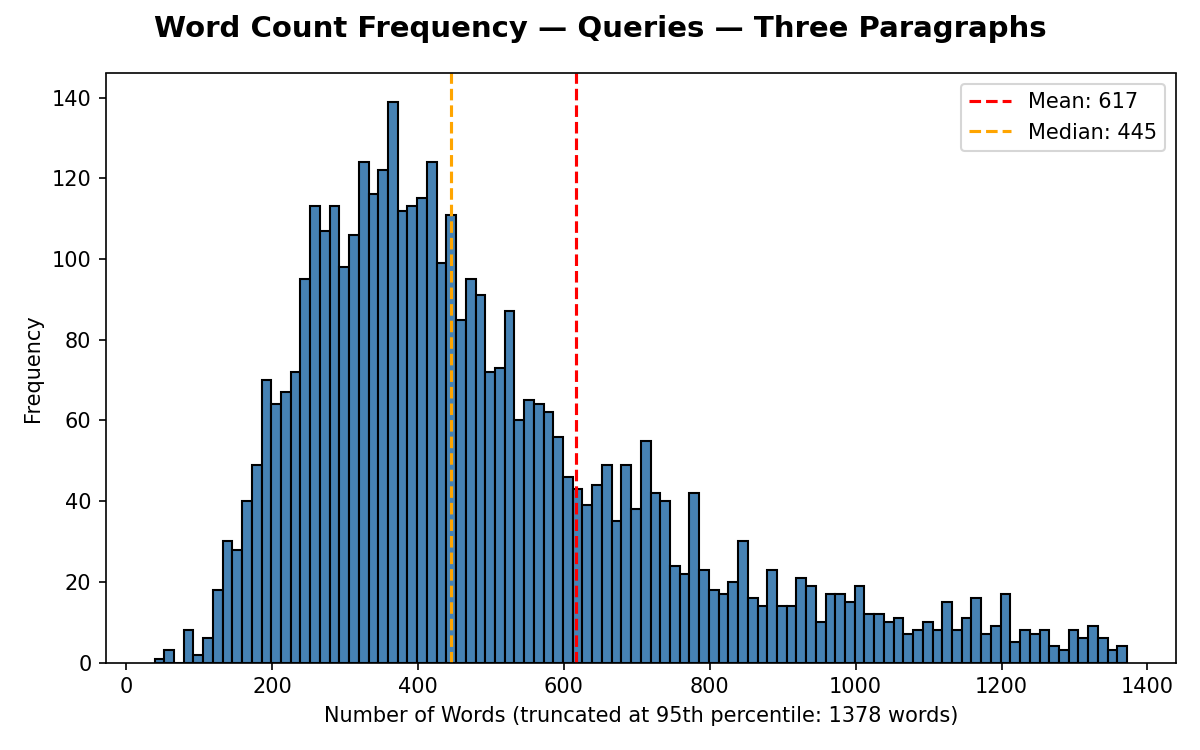}
	\caption{Word frequency graph for three paragraph queries}
	\label{fig:3p_queries}
\end{figure}

Table \ref{tab:para_stats} reports the word-count statistics for the anchor, positive, negative, and candidate texts across the benchmark splits. The corresponding distributions for the single-paragraph and contextual configurations are shown in Figures \ref{fig:1p_candidates}--\ref{fig:3p_queries}.

\begin{table}[h]
	\centering
	\resizebox{0.9\columnwidth}{!}{%
	\begin{tabular}{|l|l|r|r|r|r|}
		\toprule
		\textbf{Split} & \textbf{Role} & \textbf{Mean} & \textbf{Median} & \textbf{Min} & \textbf{Max} \\
		\midrule
		\multirow{3}{*}{Train}
		& Anchor Text   & 317.4 & 169.0 &  4 & 86{,}520 \\
		& Positive Text & 196.5 & 152.0 &  3 &  3{,}856 \\
		& Negative Text & 130.7 & 105.0 &  0 &  4{,}627 \\
		\midrule
		\multirow{3}{*}{Eval}
		& Anchor Text   & 323.1 & 167.0 & 17 & 12{,}669 \\
		& Positive Text & 206.1 & 151.5 &  3 & 18{,}791 \\
		& Negative Text & 126.2 & 105.0 &  0 &  1{,}400 \\
		\midrule
		\multirow{2}{*}{Test}
		& Query         & 327.8 & 173.0 & 14 & 25{,}750 \\
		& Positive Text & 200.6 & 154.0 &  2 & 18{,}791 \\
		\midrule
		\multicolumn{2}{|c|}{Candidates}  & 167.4 & 131.0 & 2 & 18{,}791 \\
		\bottomrule
	\end{tabular}
}
   \caption{Word-count statistics for query, positive, negative, and candidate texts across the \textsc{UK-PRBench} dataset splits, including mean, median, minimum, and maximum values.}
    \label{tab:para_stats}
\end{table}

\section{Methodology}
We formulate paragraph-level precedent retrieval as a dense retrieval problem. Given a query paragraph from a citing judgment, the objective is to rank candidate paragraphs from precedent cases such that the paragraph explicitly cited by the query is retrieved as highly as possible. We represent queries and candidate paragraphs using text-embedding models and rank candidate paragraphs according to their embedding similarity.

\subsection{Embedding Models}

We select embedding models based on their performance in two established evaluation benchmarks, the Massive Text Embedding Benchmark (MTEB) \cite{muennighoff-etal-2023-mteb} and the Massive Legal Embedding Benchmark (MLEB) \cite{butler2025massive}. We use the rankings reported by these benchmarks to identify strong-performing embedding models for our experiments. This provides a comparison between models that perform strongly on general text embedding tasks and those that perform strongly on legal-domain embedding tasks.

\subsubsection{Top-performing Models on MTEB}
We select five of the top-performing embedding models reported in MTEB, \texttt{bge\allowbreak-m3} \cite{chen-etal-2024-m3}, \texttt{nomic\allowbreak-embed\allowbreak-text\allowbreak-v1.5} \cite{nussbaum2024nomic}, \texttt{Qwen3\allowbreak-Embedding\allowbreak-0.6B} \cite{qwen3embedding}, \texttt{Qwen3\allowbreak-Embedding\allowbreak-4B} \cite{qwen3embedding}, and \texttt{snowflake\allowbreak-arctic\allowbreak-embed\allowbreak-l\allowbreak-v2.0} \cite{yu2024arctic}. These models represent strong general-purpose embedding systems according to the MTEB evaluation.

\subsubsection{Top-performing Models on MLEB}
To complement the general-purpose models, we select top-performing embedding models from the MLEB ranking \cite{butler2025massive}. MLEB evaluates embedding models specifically on legal-domain tasks, allowing us to assess whether models that perform strongly on legal embedding benchmarks are better suited to paragraph-level precedent retrieval. Specifically, we used \texttt{dinghy\allowbreak-law\allowbreak-4b\allowbreak-v1} \cite{dinghy-law-4b}, \texttt{jasper\allowbreak-token\allowbreak-compression\allowbreak-600M} \cite{zhang2025jaspertokencompression600mtechnicalreport}, and \texttt{LGAI\allowbreak-Embedding\allowbreak-Preview} \cite{choi2025lgaiembeddingpreviewtechnicalreport}.

\subsection{Experimental Settings}

We evaluate the selected embedding models under two experimental settings:

\begin{itemize}[leftmargin=0pt]
\item \textbf{Zero-shot retrieval:} In the zero-shot setting, the pretrained embedding models are used directly without any additional training on \textsc{UK-PRBench}. This setting measures how effectively existing embedding models transfer to paragraph-level legal precedent retrieval without task-specific adaptation.

\item \textbf{Fine-tuned retrieval:} In the fine-tuning setting, the embedding models described above are further trained on the training portion of \textsc{UK-PRBench} to adapt their representations to the precedent retrieval task. We use \texttt{Sentence Transformers} for fine-tuning. In addition to the embedding models described above, we include pretrained transformer models, namely \texttt{bert\allowbreak-base\allowbreak-uncased} \cite{devlin-etal-2019-bert} and \texttt{legal\allowbreak-bert\allowbreak-base\allowbreak-uncased} \cite{chalkidis-etal-2020-legal}, and fine-tune them for the retrieval task. This allows us to investigate whether task-specific fine-tuning enables smaller pretrained transformer models to perform competitively with dedicated embedding models.

The fine-tuning objective encourages the query and its cited precedent paragraph to obtain similar representations while distinguishing the positive paragraph from hard negative paragraphs originating from the same precedent case. The same fine-tuning procedure and hyperparameter configuration which are shown in Table \ref{tab:hyperparameters} are applied across all models to ensure a consistent comparison.

\end{itemize}

\subsection{Retrieval}

After model preparation, we encode all candidate paragraphs in the candidate pool into vector representations. For each query, we generate its embedding and compute cosine similarity against the candidate embeddings. We then rank candidates in descending order of similarity and return the top-$k$ candidates as the retrieval results.

For the SP configuration, each query and candidate is represented by a single paragraph. For the CP configuration, the query and candidate are represented using their corresponding three-paragraph contexts. The maximum input length is set to 4,096 tokens to accommodate longer contextual inputs.

\begin{figure}[h]
	\centering
	\includegraphics[width=0.9\columnwidth]{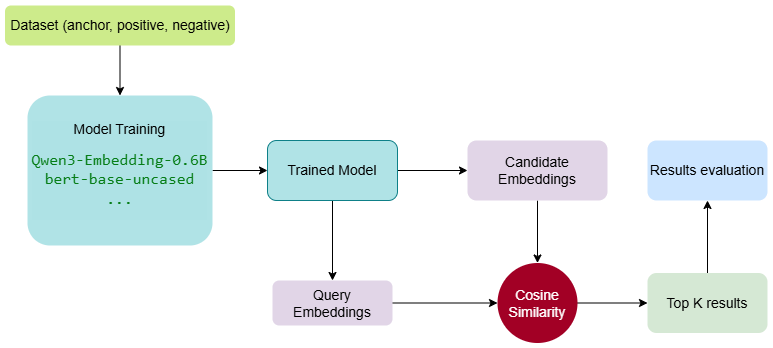}
	\caption{The paragraph retrieval modelling architecture}
	\label{fig:architecture}
\end{figure}

Figure~\ref{fig:architecture} illustrates the overall retrieval pipeline, including model preparation, embedding generation, similarity computation, and candidate ranking.

\subsection{Evaluation}

Following the previous research \cite{joshi-etal-2023-u}, we evaluate retrieval performance using recall at different values of $k$. We report results for $k$ ranging from 1 to 50, allowing us to assess both the ranking quality of the retrieved candidates and the ability of each model to retrieve the relevant precedent paragraph within increasingly larger result sets.

\begin{table}[h]
\small
	\centering
	\begin{tabular}{|l|l|}
		\hline
		\textbf{Parameter} & \textbf{Value} \\
		\hline
		Loss function         & SpladeLoss \\
		\hline
		Epochs                & 3 \\
		\hline
		Batch size (per device) & 2 \\
		\hline
		Gradient accumulation steps & 2 \\
		\hline
		Effective batch size  & 4 \\
		\hline
		Learning rate         & $2 \times 10^{-5}$ \\
		\hline
		Warmup ratio          & 0.1 \\
		\hline
		Optimiser             & AdamW (fused) \\
		\hline
		Precision             & FP16 \\
		\hline
		Max.\ input tokens          & 4{,}096 \\
		\hline
	\end{tabular}
    	
    \caption{Training hyperparameters used for fine-tuning embedding and pretrained transformer models on the \textsc{UK-PRBench} paragraph retrieval task.}
    \label{tab:hyperparameters}
\end{table}

\section{Results and Discussion}

Table~\ref{tab:results_merged} reports recall values for all models under the zero-shot and fine-tuned settings, on the single paragraph (SP) and contextual paragraph (CP) configurations. Even the best configuration retrieves the cited paragraph at rank~1 for only about half of queries, confirming that paragraph-level precedent retrieval is challenging on \textsc{UK-PRBench}. Below, we present the main findings.

\begin{table*}[t]
	\centering
	\footnotesize
	\setlength{\tabcolsep}{3pt}
	\renewcommand{\arraystretch}{1.1}
	
	\begin{tabular}{
		cl
		*{5}{>{\centering\arraybackslash}p{0.55cm}}
		|
		*{5}{>{\centering\arraybackslash}p{0.55cm}}
	}
		\toprule
		& &
		\multicolumn{5}{c|}{\textbf{SP Retrieval}} &
		\multicolumn{5}{c}{\textbf{CP Retrieval}} \\
		
		\cmidrule(lr){3-7}
		\cmidrule(lr){8-12}
		
		\textbf{Type} & \textbf{Model}
		& \mh{R@1}
		& \mh{R@5}
		& \mh{R@10}
		& \mh{R@20}
		& \mh{R@50}
		& \mh{R@1}
		& \mh{R@5}
		& \mh{R@10}
		& \mh{R@20}
		& \mh{R@50} \\
		
		\midrule
		
		\multirow{8}{*}{\textbf{Zero-shot}}
		& \texttt{bge-m3}
		& 0.33 & 0.50 & 0.57 & 0.62 & 0.68
		& 0.21 & 0.41 & 0.49 & 0.58 & 0.68 \\
		
		& \texttt{nomic-embed-text-v1.5}
		& 0.29 & 0.46 & 0.52 & 0.58 & 0.65
		& 0.14 & 0.30 & 0.38 & 0.46 & 0.58 \\
		
		& \texttt{Qwen3-Embedding-0.6B}
		& 0.36 & 0.54 & 0.62 & 0.68 & 0.76
		& 0.23 & 0.46 & 0.56 & 0.65 & 0.76 \\
		
		& \texttt{Qwen3-Embedding-4B}
		& \textbf{0.40} & 0.59 & 0.67 & 0.73 & 0.80
		& 0.27 & 0.52 & 0.61 & 0.71 & 0.81 \\
		
		& \texttt{snowflake-arctic-embed-l-v2.0}
		& 0.36 & 0.55 & 0.62 & 0.68 & 0.74
		& 0.24 & 0.47 & 0.56 & 0.66 & 0.75 \\
		
		\cmidrule(l){2-12}
		
		& \texttt{dinghy-law-4b-v1}
		& \textbf{0.40} & \textbf{0.60} & \textbf{0.68} & \textbf{0.75} & \textbf{0.81}
		& \textbf{0.30} & \textbf{0.55} & \textbf{0.65} & \textbf{0.74} & \textbf{0.84} \\
		
		& \texttt{jasper-token-compression-600M}
		& 0.35 & 0.53 & 0.61 & 0.67 & 0.75
		& 0.24 & 0.46 & 0.55 & 0.64 & 0.75 \\
		
		& \texttt{LGAI-Embedding-Preview}
		& 0.37 & 0.57 & 0.64 & 0.70 & 0.78
		& 0.23 & 0.48 & 0.57 & 0.67 & 0.78 \\
		
		\midrule
		
		\multirow{10}{*}{\textbf{Fine-tuned}}
		& \texttt{BERT}
		& 0.42 & 0.60 & 0.68 & 0.74 & 0.82
		& 0.30 & 0.50 & 0.59 & 0.67 & 0.77 \\
		
		& \texttt{Legal-BERT}
		& 0.30 & 0.50 & 0.58 & 0.66 & 0.76
		& 0.18 & 0.37 & 0.46 & 0.56 & 0.68 \\
		
		\cmidrule(l){2-12}
		
		& \texttt{bge-m3}
		& 0.14 & 0.22 & 0.26 & 0.30 & 0.36
		& 0.10 & 0.18 & 0.23 & 0.29 & 0.36 \\
		
		& \texttt{nomic-embed-text-v1.5}
		& \textbf{0.54} & \textbf{0.72} & \textbf{0.77} & \textbf{0.82} & \textbf{0.88}
		& 0.28 & 0.50 & 0.59 & 0.68 & 0.77 \\
		
		& \texttt{Qwen3-Embedding-0.6B}
		& 0.03 & 0.06 & 0.07 & 0.10 & 0.14
		& 0.01 & 0.02 & 0.04 & 0.05 & 0.09 \\
		
		& \texttt{Qwen3-Embedding-4B}
		& 0.31 & 0.50 & 0.59 & 0.67 & 0.77
		& \textbf{0.49} & \textbf{0.75} & \textbf{0.83} & \textbf{0.89} & \textbf{0.93} \\
		
		& \texttt{snowflake-arctic-embed-l-v2.0}
		& 0.11 & 0.19 & 0.22 & 0.26 & 0.33
		& 0.00 & 0.00 & 0.00 & 0.00 & 0.01 \\
		
		\cmidrule(l){2-12}
		
		& \texttt{dinghy-law-4b-v1}
		& 0.36 & 0.57 & 0.65 & 0.71 & 0.72
		& 0.46 & 0.71 & 0.79 & 0.76 & 0.92 \\
		
		& \texttt{jasper-token-compression-600M}
		& 0.37 & 0.58 & 0.66 & 0.73 & 0.80
		& 0.44 & 0.70 & 0.79 & 0.85 & 0.92 \\
		
		& \texttt{LGAI-Embedding-Preview}
		& 0.37 & 0.57 & 0.65 & 0.71 & 0.78
		& 0.23 & 0.48 & 0.57 & 0.67 & 0.78 \\
		
		\bottomrule
	\end{tabular}
	
	\caption{Paragraph-level precedent retrieval performance across zero-shot and fine-tuned models and single-paragraph and contextual retrieval settings. Recall at $k$ (R@$k$) are reported for $k \in \{1,5,10,20,50\}$.}
	\label{tab:results_merged}
\end{table*}

\begin{resultbox}
\textit{Legal-domain embedding models are among the strongest zero-shot retrievers.}
\end{resultbox}

\begin{figure*}[t]
    \centering
    \includegraphics[width=1.85\columnwidth]{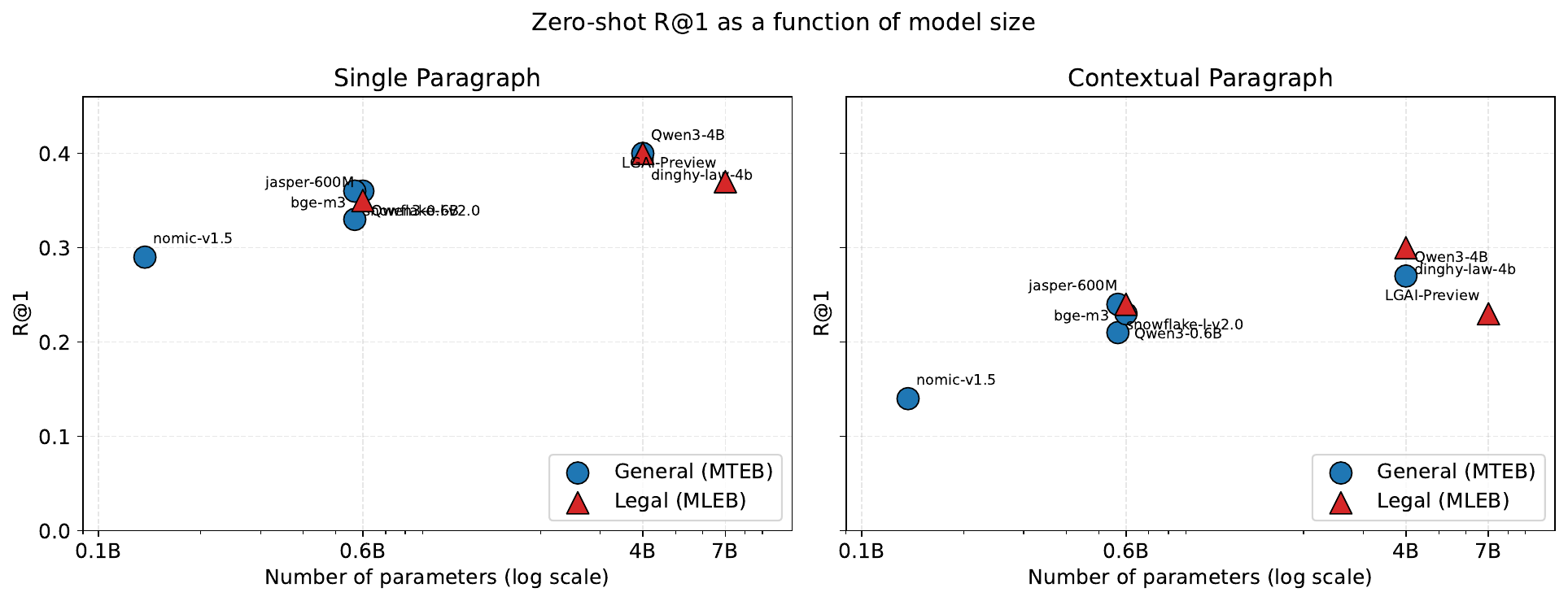}
    \caption{Zero-shot R@1 as a function of model size for general-purpose (MTEB) and legal-domain (MLEB) embedding models. Results are shown for Single-Paragraph (SP) and Contextual Paragraph (CP) retrieval.}
    \label{fig:zeroshot_r1_sp_cp}
\end{figure*}

Without task-specific fine-tuning, \texttt{dinghy-law-4b-v1} achieves the strongest overall recall performance. It obtains the highest R@1 for CP retrieval (0.30) and matches \texttt{Qwen3-Embedding-4B} for SP retrieval (0.40). It also achieves the highest recall at R@5, R@10, R@20, and R@50 in both SP and CP retrieval, reaching R@50 values of 0.81 and 0.84, respectively. This indicates that legal-domain pretraining can provide a useful advantage for precedent retrieval. However, the advantage is not universal: \texttt{Qwen3-Embedding-4B} performs comparably in SP retrieval, while several other general-purpose models remain competitive.

\begin{resultbox}
\textit{Fine-tuning produces highly model-dependent changes in retrieval performance.}
\end{resultbox}

\begin{figure}[t]
    \centering
    \includegraphics[width=\columnwidth]{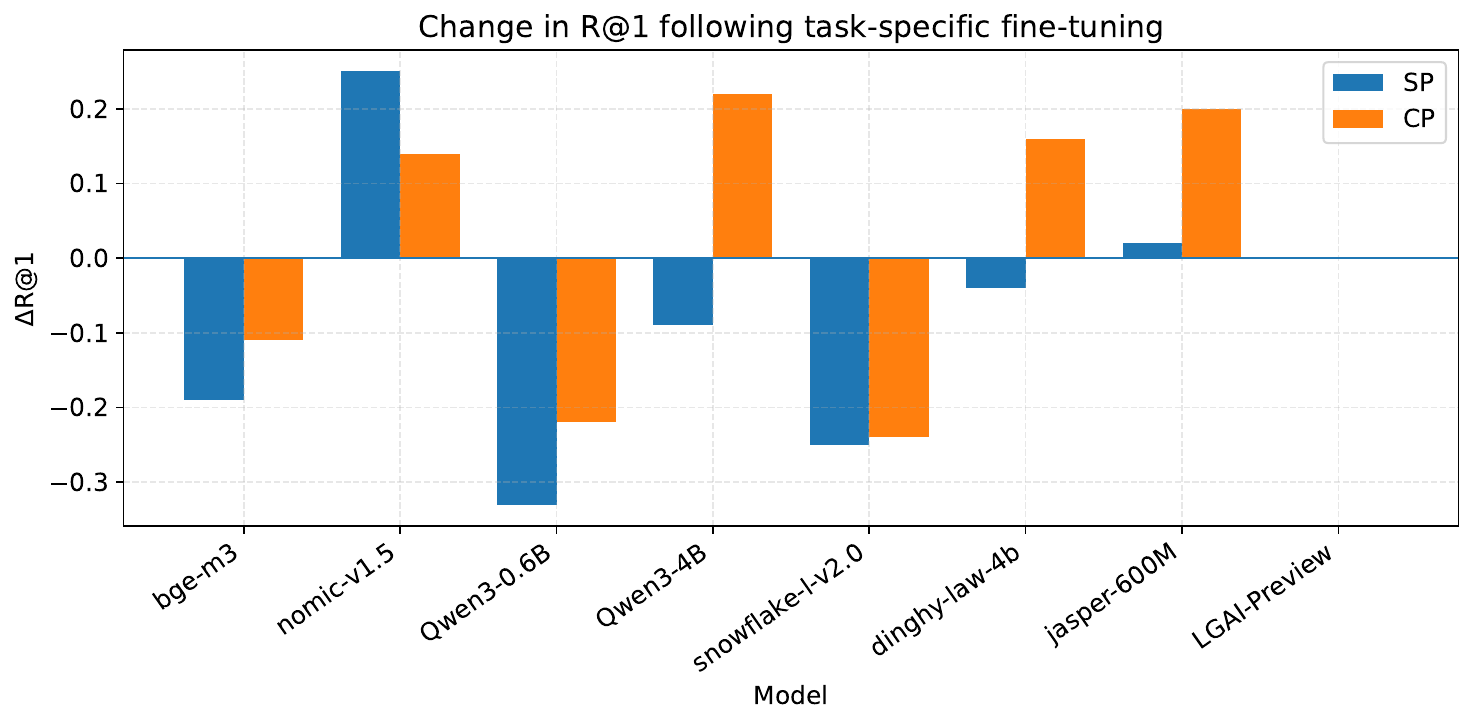}
    \caption{Change in R@1 following task-specific fine-tuning. Positive values indicate improved retrieval, while negative values indicate degraded retrieval. Results are shown for Single-Paragraph (SP) and Contextual Paragraph (CP) retrieval.}
    \label{fig:finetuning_delta_r1}
\end{figure}

The effect of task-specific fine-tuning varies substantially across embedding models. Figure~\ref{fig:finetuning_delta_r1} shows the change in R@1 following fine-tuning. For SP retrieval, \texttt{nomic-embed-text-v1.5} shows the largest improvement, increasing from 0.29 to 0.54. In contrast, fine-tuning \texttt{Qwen3-Embedding-0.6B} substantially reduces performance, with R@1 falling from 0.36 to 0.03. A similarly strong degradation is observed for \texttt{snowflake-arctic-embed-l-v2.0}, whose SP R@1 decreases from 0.36 to 0.11.

The effect is similarly heterogeneous for CP retrieval. \texttt{Qwen3-Embedding-4B} improves substantially after fine-tuning, with R@1 increasing from 0.27 to 0.49. \texttt{dinghy-law-4b-v1} also improves from 0.30 to 0.46, while \texttt{jasper-token-compression-600M} increases from 0.24 to 0.44. Conversely, \texttt{snowflake-arctic-embed-l-v2.0} falls from 0.24 to 0.00 at R@1. These results demonstrate that task-specific fine-tuning can substantially improve retrieval, but its effectiveness depends strongly on the underlying embedding model.

\begin{resultbox}
\textit{The benefit of contextualisation depends strongly on the retrieval model and training regime.}
\end{resultbox}

\begin{figure}[t]
    \centering
    \includegraphics[width=0.85\columnwidth]{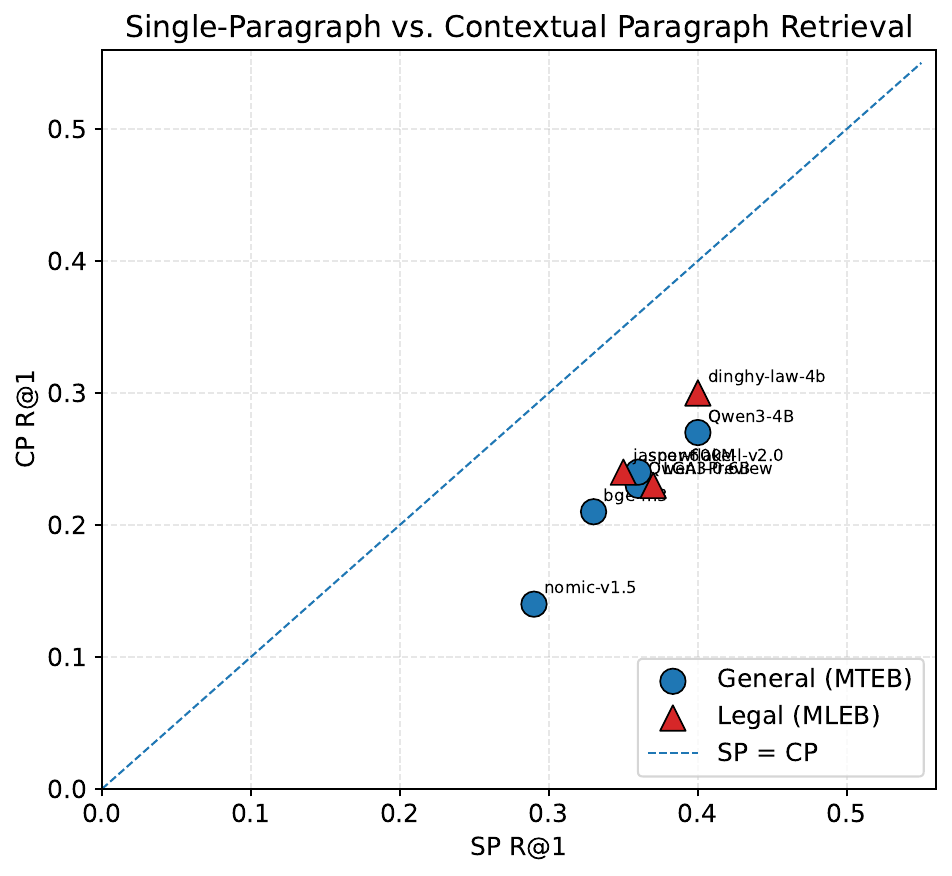}
    \caption{Relationship between Single-Paragraph (SP) and Contextual Paragraph (CP) retrieval performance measured by R@1. The dashed diagonal represents equal performance between the two settings. Points below the diagonal indicate lower R@1 with contextualisation, while points above the diagonal indicate improved R@1.}
    \label{fig:contextualisation_r1}
\end{figure}

The comparison between SP and CP retrieval shows that adding local paragraph context does not consistently improve retrieval performance. Figure~\ref{fig:contextualisation_r1} illustrates this variation using R@1. In the zero-shot setting, all models lie below the SP=CP diagonal, indicating lower R@1 when contextual information is added. For example, \texttt{nomic-embed-text-v1.5} falls from 0.29 to 0.14, while \texttt{bge-m3} decreases from 0.33 to 0.21. This suggests that, without task-specific adaptation, additional local context can make it more difficult to identify the cited paragraph at the top of the ranking.

The pattern changes after fine-tuning. Several models achieve substantially higher R@1 with CP than with SP. Most notably, \texttt{Qwen3-Embedding-4B} increases from 0.31 to 0.49, while \texttt{dinghy-law-4b-v1} increases from 0.36 to 0.46 and \texttt{jasper-token-compression-600M} from 0.37 to 0.44. However, contextualisation remains detrimental for some models, such as \texttt{nomic-embed-text-v1.5}, which decreases from 0.54 to 0.28 after fine-tuning. These results indicate that the usefulness of local context depends on both the retrieval model and its task-specific adaptation.

At deeper retrieval depths, the same pattern is observed. For example, fine-tuned \texttt{Qwen3-Embedding-4B} achieves R@50 of 0.93 with CP compared with 0.77 with SP, while \texttt{jasper-token-compression-600M} reaches 0.92 with CP compared with 0.80 with SP. Thus, contextualisation can improve both early and deeper retrieval for some models, but the effect is not consistent across systems.

\begin{resultbox}
\textit{Model scale alone is insufficient to predict paragraph-level retrieval performance.}
\end{resultbox}

Figure~\ref{fig:zeroshot_r1_sp_cp} shows that there is no consistent relationship between model size and zero-shot retrieval effectiveness. Among the Qwen models, increasing the size from 0.6B to 4B improves R@1 for both SP and CP. However, parameter count alone does not explain differences across model families. For example, the 0.137B \texttt{nomic-embed-text-v1.5} reaches a fine-tuned SP R@1 of 0.54, substantially exceeding several larger models, while the 0.6B \texttt{Qwen3-Embedding-0.6B} falls to an R@1 of only 0.03 after fine-tuning. These results suggest that architecture, pretraining, and task-specific adaptation play an important role in retrieval effectiveness beyond parameter count alone.

Overall, paragraph-level precedent retrieval remains a difficult task, particularly at the top of the ranking. The strongest zero-shot systems achieve an R@1 of 0.40, while the best fine-tuned SP system, \texttt{nomic-embed-text-v1.5}, reaches 0.54. At larger retrieval depths, performance is substantially higher, with the strongest fine-tuned models reaching R@50 values of 0.88 for SP and 0.93 for CP. The difference between R@1 and deeper recall demonstrates that many relevant paragraphs can be retrieved within the top 50 results but are not consistently ranked first. Improving the ranking of relevant paragraphs therefore remains an important challenge for paragraph-level precedent retrieval.

\section{Conclusion}

We introduced \textsc{UK-PRBench}, a benchmark for paragraph-level precedent retrieval in UK case law. Constructed from 64,284 judgments obtained from the UK National Archives, the benchmark covers a broad range of UK courts and tribunals and includes both Single-Paragraph (SP) and Contextual Paragraph (CP) retrieval settings.

Our experiments show that paragraph-level precedent retrieval remains challenging for current embedding models aligning with the recent research \cite{premasiri-etal-2023-model}. Legal-domain models perform strongly in the zero-shot setting, while fine-tuning can substantially improve performance, although its effectiveness varies across models. We also find that contextual information is not consistently beneficial, highlighting the importance of model architecture and task-specific adaptation. Overall, the relatively low R@1 scores demonstrate that accurately ranking the most relevant precedent paragraph remains an open challenge. We hope \textsc{UK-PRBench} provides a foundation for future research on fine-grained legal retrieval, including methods that better capture legal reasoning, contextual relevance, and citation relationships.

\section*{Limitations}

\textsc{UK-PRBench} reflects explicit citation behaviour rather than complete legal relevance, as uncited paragraphs may also be relevant. Its coverage is also limited by the availability of judgments in the UK National Archives collection, with some courts, periods, and cited cases unavailable. Finally, our evaluation uses a fixed candidate pool and dense paragraph retrieval, which may not fully reflect open-ended legal search involving additional signals such as citation networks, metadata, and legal reasoning.

\section*{Ethical Considerations}

We obtained a transactional licence from the UK National Archives to access the judgments used to construct \textsc{UK-PRBench}. We do not redistribute the underlying judgment texts; instead, we release the annotations and citation information. Researchers can obtain the appropriate transactional licence and use our citation extraction and instance-construction scripts to reconstruct the benchmark from the original judgments. This approach supports reproducibility while complying with the licensing conditions. As judicial records may contain sensitive information, researchers should handle the underlying documents in accordance with applicable legal, licensing, and institutional requirements.

\section*{Acknowledgments}

The experiments in this paper were conducted
in UCREL-HEX \cite{UcrelHex} and HEC \cite{pacey-hec} high-performance clusters. We
would like to thank John Vidler for the continuous
support and maintenance of the UCREL-HEX infrastructure, which enabled the efficient execution
of our experiments.

\bibliography{custom}

\end{document}